\documentclass[twocolumn,showpacs,preprintnumbers,amsmath,amssymb,superscriptaddress,nofootinbib]{revtex4-1} 
\usepackage{mathrsfs, epsfig, graphicx, url, dcolumn, bm}
\usepackage{hyperref}
\usepackage{hypernat}

\begin{document}

\title{Primordial non-Gaussianity estimation using 21 cm tomography \\ from the epoch of reionization}

\author{Yi Mao}
\email{mao@iap.fr}
\affiliation{Institut d'Astrophysique de Paris, CNRS, UPMC Univ Paris 06, UMR7095, 98 bis, boulevard Arago, F-75014, Paris, France}
\affiliation{Institut Lagrange de Paris (ILP), Sorbonne Universit\'es, 98 bis, boulevard Arago, F-75014 Paris, France}
\affiliation{Department of Astronomy and Texas Cosmology Center, University of Texas, Austin, Texas 78712, USA}
\author{Anson D'Aloisio}
\email{anson@astro.as.utexas.edu}
\affiliation{Department of Astronomy and Texas Cosmology Center, University of Texas, Austin, Texas 78712, USA}
\author{Jun Zhang}
\affiliation{Department of Physics, Shanghai Jiao Tong University, Shanghai 200240, China}
\affiliation{Department of Astronomy and Texas Cosmology Center, University of Texas, Austin, Texas 78712, USA}
\author{Paul R. Shapiro}
\affiliation{Department of Astronomy and Texas Cosmology Center, University of Texas, Austin, Texas 78712, USA}

\date{submitted 1 May 2013; accepted 11 October 2013; published 23 October 2013}

\begin{abstract}

Measuring the small primordial nonGaussianity (PNG) predicted by cosmic inflation theories may help diagnose them. The detectability of PNG by its imprint on the 21~cm power spectrum from the epoch of reionization is reassessed here in terms of $f_{NL}$, the local nonlinearity parameter. We find that an optimum, multi-frequency observation by SKA can achieve $\Delta f_{NL} \sim 3$ (comparable to recent {\it Planck} CMB limits), while a cosmic-variance-limited array of this size like {\it Omniscope} can even detect $\Delta f_{NL} \sim 0.2$. This substantially revises the methods and results of previous work. 

\end{abstract}


\setcounter{footnote}{0}


\maketitle

\section{Introduction}
The theory of cosmic inflation \cite{Guth81,Linde82}, advanced to solve the cosmological horizon and flatness problems, also explains the initial fluctuations which later gave rise to galaxies and large-scale structure in the universe. While inflation generically predicts initial matter density fluctuations with an approximately Gaussian random distribution, the small deviations from Gaussianity that characterize different inflation models have been suggested to provide an observational probe to test and distinguish the models. 
While purely Gaussian initial density fluctuations are fully described by their power spectrum, primordial non-Gaussianity (PNG) requires higher-order statistics to characterize it, the lowest-order being the 3-point correlation function, or its Fourier transform -- the bispectrum -- which is zero for the Gaussian case. Henceforth, we will describe the level of PNG predicted by different inflation models in terms of this bispectrum as parametrized by the dimensionless ``nonlinearity parameter'' $f_{NL}^{\rm local}$, specialized to the case of the so-called ``local'' template. [For further details, see \cite{D'Aloisio13} and refs.\ therein.]

The standard simplest model -- slow roll, single-field inflation -- predicts an extremely small level of PNG, given by  $f_{NL}^{\rm local}=(5/12)(1-n_\mathrm{s})$, where $n_\mathrm{s}$ is the spectral index of the primordial power spectrum \cite{Acquaviva03,Maldacena03,Creminelli04,Seery05,Chen07,Cheung08}. Recent cosmic microwave background (CMB) temperature anisotropy measurements find $n_\mathrm{s} \approx 0.96$ \cite{Planck13cosmo}, so $f_{NL}^{\rm local}\approx 0.016$. Other more general models (e.g.\ multi-field inflation) predict much larger values of $f_{NL}^{\rm local}$.

Observational cosmology has made important progress in constraining PNG thus far. Recent measurement of the CMB anisotropy bispectrum  by {\it Planck} has placed the most stringent constraint so far, $f_{NL}^{\rm local} = 2.7 \pm 5.8$  \cite{Planck13PNG}. Since, even in the ideal noise-free limit, CMB temperature (temperature+polarization) measurements can only reduce the error to $\Delta f_{NL}^{\rm local} \approx 3.5\,(1.6)$ \cite{Komatsu01,Babich04}, there is great interest in finding other methods to measure PNG; if future observations still do not detect PNG, an error budget $\Delta f_{NL}^{\rm local} < 1$ will be necessary to rule out non-standard inflationary models conclusively. [Henceforth, we focus on this ``local'' template and remove the label ``local''.]

PNG affects the clustering of the early star-forming galactic halos responsible for creating a network of ionized patches in the surrounding intergalactic medium (IGM) during the epoch of reionization (EOR), which leaves a PNG imprint on the tomographic mapping of neutral hydrogen in the IGM using its redshifted 21~cm radiation. We shall here investigate in detail the prospects for constraining PNG with radio interferometric 21~cm measurements. 
Our method and results differ significantly from previous attempts in the literature \cite{Joudaki11,Chongchitnan13}, as follows: (1) we apply the ionized density bias derived by Ref.~\cite{D'Aloisio13} to model the effect of PNG by the excursion-set model of reionization (ESMR); 
(2) we show a phenomenological model that can constrain PNG just as accurately, independent of reionization details; 
(3) we show that a single-epoch measurement can be tuned to the optimum frequency for constraining PNG; 
(4) we show how combining multi-epoch measurements further reduces the forecast errors. 

\section{PNG Signature in the 21~cm \\ power spectrum}
The 3-D power spectrum of 21~cm brightness temperature fluctuations (hereafter, ``21~cm power spectrum'') in observer's redshift space can be expressed to linear order in {\it neutral} and {\it total} hydrogen density fluctuations, $\delta_{\rho_{\rm HI}}$ and $\delta_{\rho_{\rm H}}$, respectively, as the sum of powers of $\mu_{\bf k}\equiv {\bf k} \cdot {\bf n}/|{\bf k}|$ (cosine of angle between line-of-sight (LOS) ${\bf n}$ and wave vector ${\bf k}$ of a given Fourier mode) \cite{Barkana05,Mao12}, 
$P_{\Delta T}({\bf k},z) = \widetilde{\delta T}_b^2 \bar{x}_{\rm HI}^2\,\left[ P_{\delta_{\rho_{\rm HI}},\delta_{\rho_{\rm HI}}}(k,z)  + 2\,P_{\delta_{\rho_{\rm HI}},\delta_{\rho_{\rm H}}}(k,z) \, \mu_{\bf k}^2 \right.$ $\left. + P_{\delta_{\rho_{\rm H}},\delta_{\rho_{\rm H}}}(k,z) \, \mu_{\bf k}^4 \right]$,
where $\widetilde{\delta T}_b (z) = (23.88\,{\rm mK})\left(\frac{\Omega_{\rm b}h^2}{0.02}\right)$ $\times\sqrt{\frac{0.15}{\Omega_{\rm M} h^2}\frac{1+z}{10}}$, and $\bar{x}_{\rm HI}(z)$ is the global neutral fraction. $P_{a,b}$ is the power spectrum between fields $a$ and $b$. Here, we focus on the limit where spin temperature $T_s \gg T_{\rm CMB}$, valid soon after reionization begins. As such, we can neglect the dependence on spin temperature, but our discussion can be readily generalized to finite $T_s$. We also focus on the 21~cm signal on large scales $k\le 0.15\,{\rm Mpc}^{-1}$, so that linearity conditions are met (see Ref.~\cite{Shapiro13} for a summary of these conditions). When the typical size of ionized regions is much smaller than the scale of interest, nonlinear effects of reionization patchiness on the 21~cm power spectrum \cite{Shapiro13} can be neglected. If we define {\it neutral} and {\it ionized} density biases, $b_{\rho_{\rm HI}}$ and $b_{\rho_{\rm HII}}$, according to $b_a(k) \equiv \tilde{\delta}_{a}({\bf k})/\tilde{\delta}_{\rho}({\bf k})$, i.e.\ ratio of density fluctuation in field $a$ to that of total matter density in Fourier space, 
then the 21~cm power spectrum can be rewritten as 
\begin{equation}
P_{\Delta T}({\bf k},z) = \widetilde{\delta T}_b^2 \bar{x}_{\rm HI}^2\,\left[ b_{\rho_{\rm HI}}(k,z) + \mu_{\bf k}^2 \right]^2\, P_{\delta\delta}(k,z)\,,
\label{eqn:21power}
\end{equation}
where $P_{\delta\delta}(k,z)$ is the total matter density power spectrum. Here, we assume the baryon distribution traces the cold dark matter on large scales, so $\delta_{\rho_{\rm H}} = \delta_{\rho}$. 

The ionized density bias is the fundamental quantity derived from reionization models, related to the neutral density bias by 
\begin{equation}
b_{\rho_{\rm HI}} = \left(1-\bar{x}_{\rm HII}\,b_{\rho_{\rm HII}}\right)/\bar{x}_{\rm HI}\,, 
\label{eqn:HI-HII}
\end{equation}
where $\bar{x}_{\rm HII} = 1-\bar{x}_{\rm HI}$. 
We model reionization with PNG, as follows, based on the results of Ref.~\cite{D'Aloisio13}. 

\begin{figure}[t]
\includegraphics[width=8.6cm]{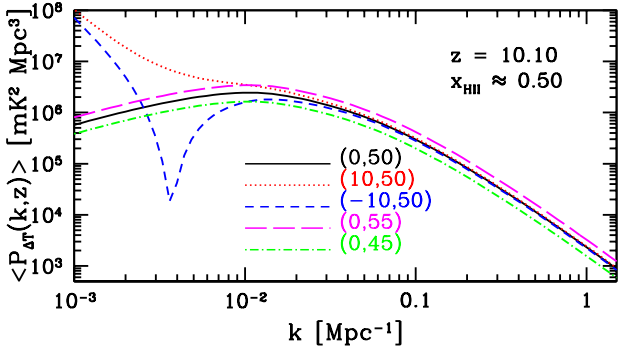}
\caption{The spherically-averaged 21~cm power spectrum at $z=10.10$, using ESMR with the values of $(f_{NL},\zeta_{\rm ESMR})$ marked in the legend.}
\label{fig:21power}
\end{figure}

(i) ESMR: 
The basic postulate of ESMR\cite{Furlanetto04} is that the local ionized fraction within a spherical volume with radius $R$ is proportional to the local collapsed fraction of mass in luminous sources above some mass threshold $M_{\rm min}$, i.e.~$x_{\rm HII}(M_{\rm min},R,z) = \zeta_{\rm ESMR}\, f_{\rm coll}(M_{\rm min},R,z)$, where $\zeta_{\rm ESMR}$ parametrizes the efficiency of this mass in releasing ionizing photons into the IGM. For simplicity, we assume atomic-cooling halos (ACHs), i.e.\  halos with virial temperature $T_{\rm vir} \gtrsim 10^4\,{\rm K}$, are the only sources of ionizing radiation. 

Our methodology for ESMR with PNG is as follows. The collapsed fraction of ACHs in Ref.~\cite{D'Aloisio12} (see also \cite{Adshead12}),  calculated in the non-Markovian extension to the excursion set formalism \cite{Maggiore10a,Maggiore10b,Maggiore10c}, for a given PNG parameter, is applied to the ESMR formalism to calculate the ionized density bias, for a given efficiency, analytically, as described in detail in Ref.~\cite{D'Aloisio13}. Henceforth, since the functions $\bar{x}_{\rm HII}(z)$ and $b_{\rho_{\rm HII}}(k,z)$ are set by two parameters $(f_{NL},\,\zeta_{\rm ESMR})$, given a fiducial cosmology, so is the 21~cm power spectrum $P_{\Delta T}({\bf k},z)$ at any $z$. 
As Figure~\ref{fig:21power} illustrates, while $\zeta_{\rm ESMR}$ changes the amplitude of the 21~cm power spectrum, $f_{NL}$ changes the shape at small $k$ significantly. 
[Note that the reionization history is virtually independent of $f_{NL}$ for $\bar{x}_{\rm HII} > 0.1$, e.g., 
for $\zeta_{\rm ESMR}=50$, $\bar{x}_{\rm HII} = 0.50$ at $z=10.105\,(10.125)$ for $f_{NL}=0\,(10)$, respectively.]
The (nonzero) minimum of the curve for $f_{NL}<0$ is at wavenumber $k_\star$, where 
$b_{\rho_{\rm HI}}(k_\star,z) \approx -1/3$. 

(ii) Phenomenological (``pheno''-) model: Just as PNG exhibits a scale-dependent effect on halo bias \cite{Dalal08,Matarrese08,Afshordi08,Desjacques11,Smith12,Adshead12,D'Aloisio12,Yokoyama12}, so we also find, in Ref.~\cite{D'Aloisio13}, a scale-dependent non-Gaussian correction to the ionized density bias, $\Delta b^{(d)}_{\rho_{\rm HII}}(k,z) = b_{\rho_{\rm HII}}(k,z) - b^{\rm G}_{\rho_{\rm HII}}(z)$, where $b^{\rm G}_{\rho_{\rm HII}}$ is the Gaussian ionized density bias and scale-independent. 
[There is also a scale-independent non-Gaussian correction, $\Delta b^{(i)}_{\rho_{\rm HII}}(z)$. However, $\Delta b^{(i)}_{\rho_{\rm HII}} \ll b^{\rm G}_{\rho_{\rm HII}}$ for $f_{NL} < 10$ (see \cite{D'Aloisio13}), so we neglect it here, similar to the neglect of a scale-independent non-Gaussian correction to the halo bias when constraining PNG with galaxy surveys \cite{Giannantonio13}.] 

For the local template, we derived from the ESMR a relation between $\Delta b^{(d)}_{\rho_{\rm HII}}$ and $b^{\rm G}_{\rho_{\rm HII}}$ in \cite{D'Aloisio13}. On large scales, 
\begin{equation}
\Delta b^{(d)}_{\rho_{\rm HII}}(k,z) = 3f_{NL} \left[ b^{\rm G}_{\rho_{\rm HII}}(z) - 1\right] 
  \frac{\delta_c \Omega_m (H_0/c)^2}{g(0) D(z)k^2 T(k)}, 
\label{eqn:sdbias}
\end{equation}
where $\delta_c \approx 1.686$ is the critical density in the spherical collapse model (in an Einstein-de Sitter universe); $D(z)$ is the linear growth factor normalized to unity at $z=0$; 
$g(0)=(1+z_i)^{-1}D^{-1}(z_i) \approx 0.76$ in our fiducial cosmology, where $z_i$ corresponds to the initial epoch, i.e.\ limit of large redshift; 
and $T(k)$ is the matter transfer function normalized to unity on large scales. 

This relation was further tested and confirmed by numerical solution of the linear perturbation theory of reionization (LPTR) which includes radiative transfer \cite{Zhang07}. [For further details, see Ref.~\cite{D'Aloisio13}.] Henceforth, Eq.~(\ref{eqn:sdbias}) is assumed to be {\it generic}, regardless of reionization details. In what we call the ``Pheno-model'', the 21~cm power spectrum is set by three parameters, $f_{NL}$, $\bar{x}_{\rm HI}$ and $b^{\rm G}_{\rho_{\rm HII}}$, at a given redshift. The latter two parameters embrace our ignorance of reionization.

\begin{table}
\caption{Specifications for 21~cm interferometers. We assume observation time $t_0 = 4000\,{\rm hours}$ for each redshift bin of bandwidth $B=6\,{\rm MHz}$.\label{tab:spec}}
\begin{ruledtabular}
\begin{tabular}{cccccc}

Experiment & $N_{\rm in}$ & $L_{\rm min}$ (m) & $\eta$ & $A_e$($z=6/8/12$)[${\rm m}^2$] & $\Omega$[sr] \footnotemark[1] \\ \hline
MWA\footnotemark[2] & 50     & 12.5 & 1   & 9/14/18	   & $\lambda^2/A_e$\\ 
LOFAR       & 32     & 100  & 0.8 & 397/656/1369 & $2(\lambda^2/A_e)$\footnotemark[3] \\
SKA         & 1400   & 10   & 0.8 & 30/50/104    & $\lambda^2/A_e$\\
Omniscope   & $10^6$ & 1    & 1   & 1/1/1        & $ 2\pi$\\
\end{tabular}
\end{ruledtabular}
\footnotetext[1]{Sky rotation adds an additional factor of two to the actual observed patches.}
\footnotetext[2]{128 antennae total \cite{mwa}.}
\footnotetext[3]{Assume LOFAR can simultaneously observe two patches on the sky.}
\end{table}

\section{Observability of 21~cm interferometric arrays using \\ Fisher matrix formalism}
Radio interferometric arrays measure the 21~cm signal from coordinates ${\bf \Theta} \equiv \theta_x \hat{e}_x +\theta_y \hat{e}_y + \Delta \nu {\bf n}$, where $(\theta_x,\theta_y)$ mark the angular location on the sky, and $\Delta \nu$ is the frequency difference from the central redshift $z_*$ of a redshift bin. It is related to the 3D Cartesian coordinates ${\bf r}$ (with origin at the bin center) by ${\bf\Theta}_\perp={\bf r}_\perp/d_A(z_*)$, and $\Delta \nu = r_\parallel / y(z_*)$, where $d_A(z)$ is the comoving angular diameter distance, $y(z) \equiv \lambda_{21}(1+z)^2/H(z)$, $\lambda_{21}= \lambda(z)/(1+z) \approx 0.21\,{\rm m}$, and $H(z)$ is the Hubble parameter at $z$. The Fourier dual of ${\bf \Theta}$ is defined as ${\bf u} \equiv u_x \hat{e}_x + u_y \hat{e}_y  + u_\parallel {\bf n}$ ($u_\parallel$ has units of time), which is related to ${\bf k}$ (Fourier dual to ${\bf r}$) by ${\bf u}_\perp = d_A {\bf k}_\perp$ and $u_\parallel = y\,k_\parallel$. The power spectrum in ${\bf u}$-space is related to that in the ${\bf k}$-space by $P_{\Delta T}({\bf u},z)= P_{\Delta T}({\bf k},z)/(d_A^2 y)$. 

For an interferometric array, a baseline ${\bf L}$ corresponds to ${\bf u}_\perp = 2\pi {\bf L}/\lambda$. (Note: Our convention is different from the observer's convention $u_\perp = L/\lambda$.) Let $\bar{n}({\bf L}_{\bf u_\perp})d^2{\bf L}$ denote the number of redundant baselines ${\bf L}_{\bf u_\perp}$ corresponding to ${\bf u_\perp}$, i.e.\ autocorrelation of array density. Then, the noise power spectrum in ${\bf u}$-space \cite{McQuinn06, Mao08} is 
$P^{N}({\bf u}_\perp)=\left( \frac{\lambda T_{\rm sys}}{A_e}\right)^2 /\left[t_0 {\bar n}({\bf L}_{{\bf u}_\perp})\right]$, 
where $T_{\rm sys} \approx (280\,{\rm K})\left[(1+z)/7.4\right]^{2.3}$ is the system temperature \cite{Wyithe07}, $A_e \propto \lambda^2$ (for $\lambda$ less than antenna size) is effective collecting area, and $t_0$ is total observation time.

We adopt the following configuration of interferometric arrays. We assume antennas are concentrated within a {\it nucleus} of radius $R_0$ with area coverage fraction close to 100\%, with coverage density dropping like $r^{-2}$ in a {\it core} extending from $R_0$ to $R_{\rm in}$. We neglect the dilute antenna distribution in the outskirts, $R>R_{\rm in}$. Given central array density $\rho_0$, the configuration can be specified by two convenient parameters: $N_{\rm in}$, the number of antennas within $R_{\rm in}$, and $\eta$, the fraction of these antennas that are in the nucleus . The relations are $R_0 =  \sqrt{\eta N_{\rm in}/\rho_0 \pi}$, $R_{\rm in} = R_0 \exp{[(1-\eta)/(2\eta)]}$ \cite{Mao08}. 

Given a parameter space $\{p_a\}$, the Fisher matrix for 21~cm power spectrum measurements is 
${\bf F}_{ab}=\sum_{\bf u} \left(\frac{\partial P_{\Delta T}({\bf u})}{\partial p_a}\right) \left(\frac{\partial P_{\Delta T}({\bf u})}{\partial p_b}\right)/\left[\delta P_{\Delta T}({\bf u})\right]^2$. 
The 1$\sigma$ forecast error of the parameter $p_a$ is given by $\Delta p_a = \sqrt{({\bf F^{-1}})_{aa}}$. 
The power spectrum measurement error in a pixel at ${\bf u}$ is 
$\delta P_{\Delta T}({\bf u})=  \left[P_{\Delta T}({\bf u})+P_N(u_\perp)\right]/\sqrt{N_c}$, where $N_c=u_\perp du_\perp du_\parallel \Omega B/(2\pi)^2$ is the number of independent modes in that pixel. We adopt logarithmic pixelization, $du_\perp/u_\perp = du_\parallel/u_\parallel = 10\%$. Here, $\Omega$ is solid angle spanning the field of view, $B$ is bandwidth of the redshift bin. 

We assume experimental specifications in Table~\ref{tab:spec}, for MWA\cite{mwa}, LOFAR\cite{lofar}, SKA\cite{SKA}, and Omniscope\cite{FFTT}, respectively.
(We note that these interferometer array configurations were not designed to optimize the experiment proposed here, so improved constraints may be possible with other designs.) 
We assume residual foregrounds can be neglected for $k_\parallel \ge k_{\parallel,{\rm min}}= 2\pi/(yB)$, 
e.g., $k_{\parallel,{\rm min}}= 0.055\,{\rm Mpc}^{-1}$ at $z=10.1$. This foreground removal requirement is achievable, as demonstrated by, e.g.,\cite{McQuinn06}. 
(Ref.~\cite{Lidz13}, submitted at about the same time as our paper, considers a somewhat more pessimistic foreground removal scenario in constraining PNG, complementary to our work.)
The minimum $k_\perp$ is set by the minimum baseline, $k_{\perp,{\rm min}} = 2\pi L_{\rm min}/(\lambda d_A)$. We account for modes up to $k_{\rm max} = 0.15\,{\rm Mpc}^{-1}$. Our fiducial cosmology is as follows: $\Omega_\Lambda=0.72$, $\Omega_{\rm M}=0.28$, $\Omega_{\rm b}=0.046$, $H_0 = 100h\,{\rm km}\,{\rm s}^{-1}\,{\rm Mpc}^{-1}$ ($h=0.7$), $\sigma_8=0.82$, $n_\mathrm{s}=0.96$, consistent with {\it WMAP7} results \cite{Komatsu11}, with linear matter power spectrum of Ref.~\cite{Eisenstein99}. ESMR fiducial values are $f_{NL}=0$, $\zeta_{\rm ESMR}=50.0$, corresponding to electron scattering optical depth $\tau_{\rm es}=0.08$. 
To facilitate direct comparison with ESMR, the fiducial model of 21~cm power spectrum for the Pheno-model will be the same as the ESMR fiducial model, which sets the fiducial values of $\bar{x}_{\rm HI}$ and $b^{\rm G}_{\rho_{\rm HII}}$ for a given redshift. 

\begin{table}
\caption{Forecast 1$\sigma$ errors at $\bar{x}_{\rm HII} \approx 0.25$($z=11.24$) and $\bar{x}_{\rm HII} \approx 0.50$ ($z=10.10$), respectively. ``F.V.'' means fiducial values. }
\begin{ruledtabular}
\begin{tabular}{cccccccc}
	 &   &   &  \multicolumn{2}{c}{ESMR}         &   \multicolumn{3}{c}{Pheno-model}	                \\
	      \cline{4-5}\cline{6-8} 
$\bar{x}_{\rm HII}$  &   &   & $f_{NL}$ & $\zeta_{\rm ESMR}$ & $f_{NL}$ & $\bar{x}_{\rm HI}$ & $b^{\rm G}_{\rho_{\rm HII}}$ \\
	      \hline
     & &  $\bigl\lbrack$F.V.    & 0	& 50.0   & 0        & 0.75     & 6.19 $\bigr\rbrack$  	     \\ 
        & MWA	     & &    13000    & 1500    &   14000    & 300	   & 8800		\\
$0.25$  & LOFAR	     & &    1200     & 130     &   1200     & 1.1	   & 29 		\\
        & SKA	     & &    16       & 1.8     &   16	    & 0.028	   & 0.79		\\
        & Omniscope  & &    0.38     & 0.040   &   0.38     & 0.00044	   & 0.012		\\ \hline
     & & $\bigl\lbrack$F.V.    & 0	     & 50.0	 &  0	     & 0.50 & 5.43 $\bigr\rbrack$ 	    \\ 
        & MWA	     & &      700      & 63    &  750	 & 17		& 220	 \\
$0.50$  & LOFAR      & &      100      & 8.1   &  96	 & 0.16 	& 2.0	 \\
        & SKA	     & &      19       & 1.5   &  18	 & 0.030	& 0.37   \\
        & Omniscope  & &      1.8      & 0.15  &  1.8	 & 0.0023	& 0.027  \\
	
\end{tabular}
\end{ruledtabular}
\label{tab:1sig25}
\end{table}

\section{Results}

(i) Single epoch constraints: 
In Table~\ref{tab:1sig25}, we list forecast errors, $\Delta f_{NL}$, marginalized over $\zeta_{\rm ESMR}$ in the ESMR, and over $\bar{x}_{\rm HI}$ and $b^{\rm G}_{\rho_{\rm HII}}$ in the Pheno-model.  Top and bottom sets of forecasts use information from a single redshift bin centered at $z=11.24$ ($\bar{x}_{\rm HII} = 0.25$) and $z=10.10$ ($\bar{x}_{\rm HII} = 0.50$), respectively. For the same redshift and experiment, the values of $\Delta f_{NL}$ match very well between these two models. This demonstrates that the Pheno-model, which makes no assumptions about the connection between $\bar{x}_{\rm HI}$ and $b^{\rm G}_{\rho_{\rm HII}}$, can constrain $f_{NL}$ for a single epoch measurement as accurately as a reionization model which links the evolutions of $\bar{x}_{\rm HI}$ and $b^{\rm G}_{\rho_{\rm HII}}$. 

\begin{figure}[t]
\includegraphics[width=8.6cm]{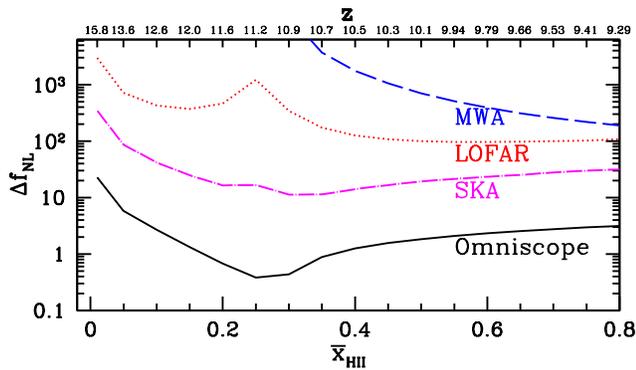}
\caption{1$\sigma$ error $\Delta f_{NL}$ from a series of single redshift bins, each with 6~MHz bandwidth.}
\label{fig:redvar}
\end{figure}

For the same experiment, as Table~\ref{tab:1sig25} shows, values of $\Delta f_{NL}$ differ significantly between $\bar{x}_{\rm HII} = 0.25$ and $0.50$, implying a strong dependence of $\Delta f_{NL}$ on $\bar{x}_{\rm HII}$. For Omniscope, in particular, $\Delta f_{NL}$ shrinks by a factor of $\sim 5$ from $\bar{x}_{\rm HII} = 0.50$ to $0.25$. To investigate this in detail, we use the ESMR to plot $\Delta f_{NL}$ vs $\bar{x}_{\rm HII}$ in Figure~\ref{fig:redvar}.  For MWA and LOFAR (both noise-dominated experiments), the constraint is tighter at higher $\bar{x}_{\rm HII}$ (i.e. lower redshift, where the noise is smaller). However, for SKA and Omniscope (cosmic-variance-dominated experiments), there appears to be a ``sweet spot,'' where $\Delta f_{NL}$ is minimized, at $\bar{x}_{\rm HII} \approx 0.25 - 0.30$. 
This sweet spot can be explained using the Pheno-model, as follows. The derivative $\left|dP_{\Delta T}/df_{NL}\right| \propto \left(\bar{x}_{\rm HII}/\bar{x}_{\rm HI}\right)\left[ b^{\rm G}_{\rho_{\rm HII}} - 1\right]\left| b^{\rm G}_{\rho_{\rm HI}} + \mu_{\bf k}^2\right|
\propto \left|(1-b^{\rm G}_{\rho_{\rm HI}})(\frac{1}{3}+b^{\rm G}_{\rho_{\rm HI}})\right|$, using Eq.~(\ref{eqn:HI-HII}). In the ideal noise-free limit, the power spectrum error $\Delta P_{\Delta T}\propto P_{\Delta T} \propto \left[ b^{\rm G}_{\rho_{\rm HI}} + \mu_{\bf k}^2\right]^2 \approx {b^{\rm G}_{\rho_{\rm HI}}}^2 + \frac{2}{3}\,b^{\rm G}_{\rho_{\rm HI}}+\frac{1}{5}$, when averaged over $\mu_{\bf k}$. 
Fixing $\bar{x}_{\rm HI}$ and $b^{\rm G}_{\rho_{\rm HII}}$, the Fisher matrix  ${\bf F} = (\Delta f_{NL})^{-2}$ 
is peaked when $b^{\rm G}_{\rho_{\rm HI}} \approx -0.71$. 
Eq.~(\ref{eqn:HI-HII}) gives $\bar{x}_{\rm HII} = (1-b^{\rm G}_{\rho_{\rm HI}})/(b^{\rm G}_{\rho_{\rm HII}}-b^{\rm G}_{\rho_{\rm HI}})$. Since Gaussian bias $b^{\rm G}_{\rho_{\rm HII}} \approx 6$ (see Fig.5 of Ref.\cite{D'Aloisio13}), the sweet spot is at $\bar{x}_{\rm HII} \approx 0.25$ in the ideal noise-free limit. In reality, finite noise is larger at higher redshifts, so the sweet spot will occur at slightly lower redshift.

\begin{table}
\caption{$\Delta f_{NL}$ from multiple redshift bins, marginalized over $\zeta_{\rm ESMR}$.
``1-band,'' ``3-band,'' ``5-band,'' ``7-band'' means the information of 1 bin (at $z = 11.24$, $\bar{x}_{\rm HII} = 0.25$), 3 bins ($\bar{x}_{\rm HII} \approx 0.17 - 0.37$), 5 bins ($\bar{x}_{\rm HII} \approx 0.11 - 0.52$), 7 bins ($\bar{x}_{\rm HII} \approx 0.06 - 0.70$), respectively.
}
\begin{ruledtabular}
\begin{tabular}{cccccc}
 Experiment    &     1-band	&   3-band    &    5-band    &    7-band   \\ \hline
 MWA	       &     13000	& 1800        & 520	     & 200	 \\
 LOFAR         &     1215	& 91	      & 39	     & 26	 \\
 SKA	       &     16 	& 5.0	      & 3.5	     & 2.8	 \\
 Omniscope     &     0.38	& 0.23        & 0.18	     & 0.16	 \\	     
\end{tabular}
\end{ruledtabular}
\label{tab:1sigall}
\end{table}

(ii) Multi-epoch constraints: 
While a futuristic single epoch measurement can achieve a remarkable accuracy of $\Delta f_{NL} = 0.38$ (16) for Omniscope (SKA), adding tomographic information can further improve the accuracy. 
The Pheno-model alone cannot be used to combine multi-frequency measurements because it does not specify the redshift evolutions of $\bar{x}_{\rm HI}$ and $b^{\rm G}_{\rho_{\rm HII}}$.  On the other hand, a model such as the ESMR can be used to combine multi-frequency measurements because it fixes the reionization history (and therefore $P_{\Delta T}$) for a given ($f_{NL}$, $\zeta_{\rm ESMR}$). 
We show multi-epoch constraints from the ESMR in Table~\ref{tab:1sigall}. 
Specifically, if information is combined from $\bar{x}_{\rm HII} \approx 0.06 - 0.70$ (7-band, total 42~MHz bandwidth, 
corresponding to $z \approx 9.5 - 13.4$ in the ESMR), the constraint can be significantly tightened, i.e.~$\Delta f_{NL} = 0.16$ for Omniscope (10~times smaller than an ideal CMB experiment), and $\Delta f_{NL} = 2.8$ by SKA (two times smaller than {\it Planck}). A prior of $\Delta\tau_{\rm es}=0.014$ from {\it Planck}+{\it WMAP} CMB measurements\cite{Planck13cosmo} corresponds to a prior of $\Delta\zeta_{\rm ESMR}\approx 39$, much larger than allowed by SKA and Omniscope alone, so adding this $\tau_{\rm es}$ prior cannot improve $\Delta f_{NL}$ from these experiments. 
If we take a more conservative upper limit, $k_{\rm max} = 0.10\,{\rm Mpc}^{-1}$ (instead of $0.15\,{\rm Mpc}^{-1}$ as assumed above), then $\Delta f_{NL}$ for Omniscope is $\sim 2$ times larger. 
Since multi-epoch observations tighten $f_{NL}$ constraints by combining information from different frequency bands to increase the amount of data relative to a single band, our use of the simple ESMR model here, with constant efficiency parameter $\zeta_{\rm ESMR}$, for which reionization spans a relatively narrow range of redshift, may be a conservative one.  If reionization is more extended, as in self-regulated reionization models\cite{Iliev07,Iliev12}, for example, the resulting $f_{NL}$ constraints may be even tighter.

\section{Comparison with previous work} 
Previously, Ref.~\cite{Joudaki11} reported forecasts of $\Delta f_{NL} = [100,700,50,4]$ for [MWA512, LOFAR, SKA, Omniscope] based on information from a single redshift bin at the 50\%-ionized epoch. Their results can be compared to the bottom set in Table \ref{tab:1sig25}.  Our results differ for a number of reasons: (1) Ref.\cite{Joudaki11} computed the scale-dependent signature of PNG in $P_{\Delta T}$ by fitting their approximate semi-numerical simulations of reionization.  In Ref.\cite{D'Aloisio13}, we showed by our analytical derivation and numerical LPTR calculations that this fit underestimates the scale-dependent bias due to PNG significantly.  These differences are reflected in the smaller $\Delta f_{NL}$ we obtain for LOFAR, SKA, and Omniscope. (2) The anticipated MWA512 configuration assumed by Ref.~\cite{Joudaki11} was also overly optimistic, while we adopt the current MWA128 configuration \cite{mwa}.  

To model the effect of PNG on the 21~cm power spectrum, Ref.~\cite{Chongchitnan13} assumed the simple functional dependence of ionized fraction on local overdensity, used for illustrative purposes by Ref.~\cite{Alvarez06} for the Gaussian case, to derive an ionized fraction bias $b_x$ for PNG. Unfortunately, they incorrectly used $b_x$ to relate the 21~cm power spectrum to the matter power spectrum (see Ref.~\cite{D'Aloisio13}). 

\section{Conclusions}
This paper suggests two approaches to constrain $f_{NL}$ with 21~cm power spectra from the EOR. If we take a conservative approach, i.e.\ assuming nothing about $\bar{x}_{\rm HI}(z)$ and $b_{\rho_{\rm HII}}^G(z)$, then the Pheno-model can be employed for a single redshift bin to provide $f_{NL}$ constraints with the same precision as a reionization model in which the reionization history is uniquely specified by a set of model parameters. However, using the ESMR, we demonstrate that a well-motivated reionization model can improve $\Delta f_{NL}$ in two ways: (1) a pathfinder measurement at a single redshift can best-fit the values of model parameters, which then can be used to estimate the desired redshift corresponding to $\bar{x}_{\rm HII} \approx 0.25 - 0.3$, i.e. the ``sweet spot'' for cosmic-variance-dominated experiments. This can help tune {\it single-epoch} observations for maximum precision. (2) {\it Multi-epoch} measurements can be combined to improve $\Delta f_{NL}$. We find that multi-frequency observation by SKA can achieve $\Delta f_{NL} \sim 3$, providing a new method to constrain PNG independent of CMB measurements, but with a precision comparable to {\it Planck}'s. A cosmic-variance-limited array of this size like {\it Omniscope} can achieve $\Delta f_{NL} \sim 0.2$, improving current constraints by an order of magnitude. 
These high precision observations may someday shed light on inflationary models.

\begin{acknowledgments}
We thank Shahab Joudaki and Mario Santos for additional information on their work in Ref.~\cite{Joudaki11}. This work was supported by French state funds managed by the ANR within the Investissements d'Avenir programme under reference No.~ANR-11-IDEX-0004-02, by U.S.~NSF Grants No.~AST-0708176 and No.~AST-1009799, and NASA Grants No.~NNX07AH09G and No.~NNX11AE09G. 
\end{acknowledgments}

\end{document}